\documentclass[prd,floatfix, nofootinbib,showpacs,
preprintnumbers
]{revtex4}
\usepackage{amsmath}
\usepackage{amssymb}
\usepackage{epsfig}
\usepackage{graphicx}
\usepackage{color}
\usepackage{hyperref}
\begin{document}

\title{Theoretical Uncertainties in Proton Lifetime Estimates}

\author{Michal Malinsk\'{y}\footnote{Presenting author; email:malinsky@ipnp.troja.mff.cuni.cz}\hskip 2mm}
\affiliation{Institute of Particle and Nuclear Physics,
Faculty of Mathematics and Physics,
Charles University in Prague, V Hole\v{s}ovi\v{c}k\'ach 2,
180 00 Praha 8, Czech Republic
}

\author{Timon Mede}\affiliation{Institute of Particle and Nuclear Physics,
Faculty of Mathematics and Physics,
Charles University in Prague, V Hole\v{s}ovi\v{c}k\'ach 2,
180 00 Praha 8, Czech Republic
}

\author{Helena Kole\v{s}ov\'a}\affiliation{Institute of Particle and Nuclear Physics,
Faculty of Mathematics and Physics,
Charles University in Prague, V Hole\v{s}ovi\v{c}k\'ach 2,
180 00 Praha 8, Czech Republic \\ and \\Faculty of Nuclear Sciences and Physical Engineering, Czech Technical University in Prague, B\v{r}ehov\'a 7, \\ 115 19 Praha 1, Czech Republic
}

\begin{abstract}
We recapitulate the primary sources of theoretical uncertainties in proton lifetime estimates in renormalizable, four-dimensional \& non-supersymmetric grand unifications that represent the most conservative framework in which this question may be addressed at the perturbative level. We point out that many of these uncertainties are so severe and often even irreducible that there are only very few scenarios in which an NLO approach, as  crucial as it is for a real testability of any specific model, is actually sensible. Among these, the most promising seems to be the minimal renormalizable SO(10) GUT whose high-energy gauge symmetry is spontaneously broken by the adjoint and the five-index antisymmetric irreducible representations. 
\end{abstract}

\maketitle
\section{INTRODUCTION}
The continuous efforts to bring into life a new generation of the large neutrino detectors such as DUNE~\cite{Adams:2013qkq} (a 40kt LAr TPC detector to be built in the Homestake mine in South Dakota), Hyper-K~\cite{Abe:2011ts} (about 500kt fiducial water-Cherenkov detector proposed in order to supersede the ``smaller'' Super-K~\cite{Nishino:2012ipa} in Japan) or perhaps even some variant of the liquid-scintillator machine like the european LENA bring back the questions of the possible fundamental instability of the baryonic matter. 
It is namely the close complementarity of the planned super-rich neutrino physics programme (with a common belief that not only the neutrino mass hierarchy would be determined but also signals of CP violation in the lepton sector may be observed) with the nucleon decay searches that fuels the hope that processes like $p\to \pi^{0}e^{+}$ or $p\to \pi^{+}\overline\nu$ (or perhaps even $p\to K^{+}\overline\nu$ favoured by low-energy supersymmetry) may finally be seen.
To this end, the projected sensitivity of these new facilities should in practically all channels exceed that of the previous generation (dominated by Super-K) by at least one order of magnitude, thus touching the ``psychological'' proton lifetime boundary of $10^{35}$ years. 

Unfortunately, the steady progress on the experimental side has hardly been complemented by any significant improvement in the accuracy of the proton lifetime  predictions in theory. This, of course, would not be a true concern on day 1 after the proton decay discovery; however, whenever it would come to more delicate (i.e., quantitative) questions beyond the obvious ``Is proton absolutely stable?'' like, e.g., ``Which models can now be ruled-out?'' or ``What did we really learn about the (presumably) unified dynamics behind these processes?'' the theory would have a difficult time to pull any robust answer up the sleeve.

This has to do, namely, with the enormous (and often irreducible) theoretical uncertainties plaguing virtually all proton lifetime estimates in the current literature\footnote{See, e.g., ~\cite{Nath:2006ut,Babu:2013jba} and the references therein.} which by far (often by many orders of magnitude) exceed the relatively small -- yet fantastic -- factor-of-ten ``improvement window'' the new generation of facilities may open. The typical reason behind this is either a disparity between the amount of the input information available and needed for any potentially accurate calculation -- with supersymmetric grand unified theories (SUSY GUTs) as a canonical example --  or the lack of a fully consistent treatment at better than just the leading order (LO). While the first issue is more a matter of fashion and, as such, it can be expected to resolve by itself in the future (especially if the LHC sees no hints of SUSY at the TeV scale) the latter is much more difficult to deal with\footnote{For some very recent attempts to get robust proton lifetime estimates in the realm of non-SUSY SO(10) GUTs see, e.g.,~\cite{Bertolini:2013vta,Kolesova:2014mfa,Babu:2015bna}.} due to the parametrically higher level of complication such a next-to-leading (NLO) analysis represents.

In this contribution we will recapitulate the main sources of theoretical uncertainties one has to take into account in any attempt to provide any such NLO proton lifetime prediction focusing on some of the existing options to overcome at least the most pressing of the relevant issues. On this ground we shall attempt to justify our belief that perhaps the only perturbative and renormalizable grand unified scenario in which a robust and consistent NLO proton lifetime calculation may be done is the non-supersymmetric SO(10) grand unified model~\cite{Chang:1984qr,Bertolini:2009es} in which the GUT-scale gauge symmetry is spontaneously broken by the adjoint scalar representation. 
  
\section{THEORETICAL UNCERTAINTIES IN PROTON LIFETIME ESTIMATES}
\begin{figure}[h]
\centerline{a) \includegraphics[width=100pt]{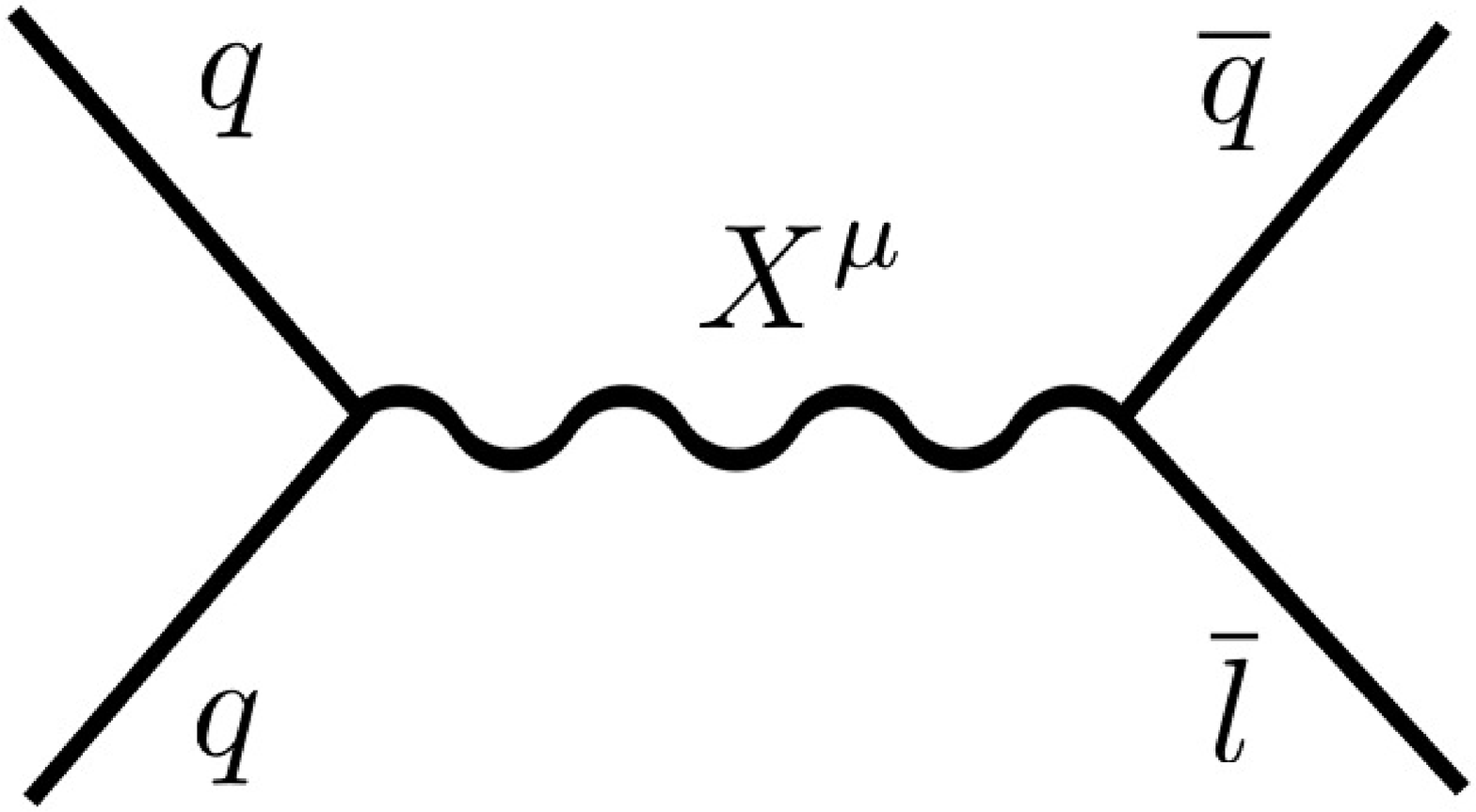}\hskip 1cm
b) \includegraphics[width=100pt]{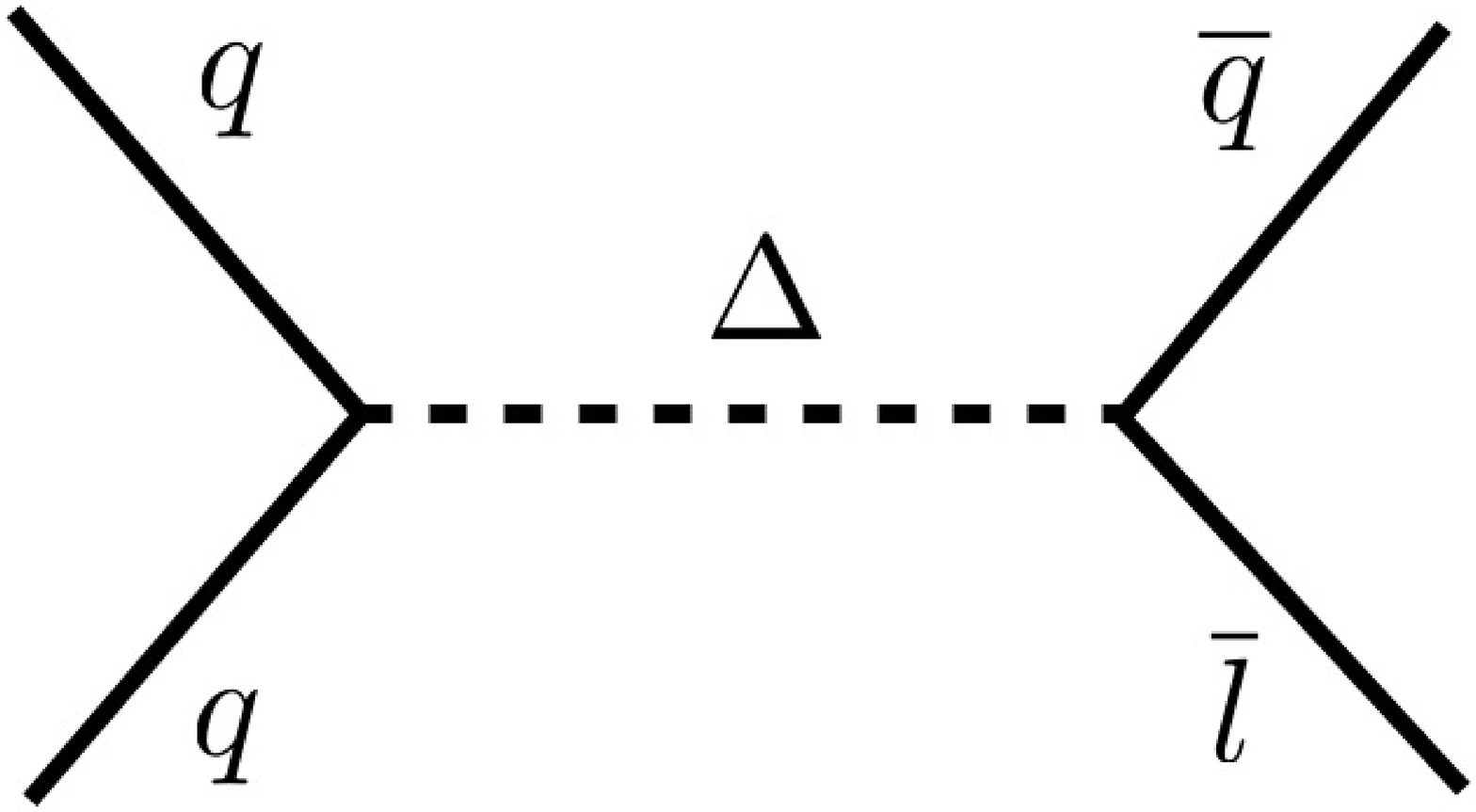}}
\caption{\label{d6operator}The typical ``internal structure'' of the gauge-mediated (left) and scalar-mediated (right) $d=6$ baryon and lepton number violating operators in the Standard model. The $X^{\mu}$ stands for the vector leptoquarks transforming, usually, as $(\overline 3,2,+\frac{5}{6})+h.c.$ or  $(\overline 3,2,-\frac{1}{6})+h.c.$ under the SM gauge group while $\Delta$ is a generic symbol for scalar mediators transforming typically as $(\overline 3,1,+\frac{1}{3})$ or  $(\overline 3,1,+\frac{4}{3})$.}
\end{figure}

There are actually many sources of theoretical uncertainties of very different origins that have to be taken into account when the ambition is to get the total uncertainty at least within the ballpark of the aforementioned experimental ``improvement window''. In the framework of the classical GUTs, these are, traditionally, i) the limited accuracy of the existing estimates of the relevant  hadronic matrix elements, ii) the limited accuracy of the determination of the masses of the $d=6$ baryon and lepton number violating (BLNV) operators (cf. FIG.~\ref{d6operator}) and iii) the insufficiency of the information about their flavour structure available at low energy. On top of that, the proximity of the GUT scale $M_{G}$, typically in the $10^{16}$ GeV ballpark, to the (reduced) Planck scale $M_{Pl}\sim 10^{18}$ GeV should make one worried about the size of the possible gravity-induced effects which, typically, are not under a good control. 
 
\subsection{Hadronic matrix elements}
Concerning the hadronic matrix elements of the (quark part of the) $d=6$ BLNV operators that provide the translation between the ``hard'' processes at the quark level and the destruction/production of real hadrons assumed in the initial and final states the current accuracy of their lattice/$\chi$PT determination can be appreciated in, e.g., Table III of the recent study~\cite{Aoki:2013yxa}. Remarkably enough, the typical uncertainties quoted there do not exceed few tens of percent (to be compared with the order-of-magnitude variations in the early days' calculations, cf.~\cite{Hara:1986hk}). Hence, the steady progress in this field has gradually tamed the errors in the hadronic matrix elements of the BLNV operators to such a degree that, at the desired one-order-of-magnitude accuracy level, one does not need to worry about them much anymore.   

\subsection{The GUT-scale effective mediator mass determination}
A way more critical class of quantities entering all the proton lifetime estimates in GUTs are the masses of the vector and scalar mediator fields in the unified-theory Feynman graphs underpinning the effective BLNV $d=6$ operators in the SM like those in FIG.~\ref{d6operator} because, unlike the hadronic elements, these enter the relevant widths as inverse fourth power. Moreover, as they are determined from the gauge unification constraints\footnote{Let us note that, in the simplest SU(5) GUT at one loop, the masses of the $(\overline 3,2,+\frac{5}{6})+h.c.$ gauge bosons correspond to the energy at which the running SU(3), SU(2) and U(1) couplings meet, i.e., the ``natural'' matching scale in the $\overline{\rm MS}$ scheme; however, this is also the only case when the situation is as simple as this. In more complicated cases (e.g., going to two loops, having more than just a single heavy vector at play etc.) the matching scale has no obvious meaning and the heavy gauge masses can be derived only when a fully consistent picture including, among other things, threshold corrections, is attained.} imposed on the running couplings which scale with energy only logarithmically, even small theoretical uncertainties in the entire picture can be hugely magnified in the lifetime estimates. 
\subsubsection{Perturbation theory uncertainties: one-loop vs. two-loop running}
To this end, one may easily quantify the uncertainties in the $M_{G}$ determination due to, e.g., omitting higher order effects in the gauge beta functions. For a rough picture it is sufficient to compare the positions (i.e., energies $\mu^{1-loop}$, $\mu^{2-loop}$) of the intersection points of two linear curves corresponding to the $\alpha^{-1}$ evolution emanating from the measured values of the gauge couplings at the low scale $\alpha_{1,2}^{-1}$ in two slightly different directions (defined by their one and two-loop $\beta$-function coefficients $b\equiv 16\pi^{2} \beta$, respectively). A simple calculation yields
\begin{equation}
\mu^{2-loop}/\mu^{1-loop}=\exp(2\pi k t)\label{effect}
\end{equation}
where $t\equiv 1/2\pi \log \mu/m_{Z}$ is the usual ``log-of-energy'' variable (for $\mu=M_{G}\sim10^{16}$ GeV one has $t\sim 5.1$) and $k$ is the ratio between  the typical size of the one- and two-loop beta functions (actually, $k$ is defined as $b^{2-loop}\sim k (b_{2}-b_{1})^{1-loop}$ which, in practice, is not just $1/(16\pi^{2})$ but typically much larger, especially if the two curves tend to meet under a ``shallow angle'', i.e., for $|(b_{2}-b_{1})^{1-loop}|\ll |b_{1,2}^{1-loop}|$. In terms of the widths this yields 
$
\Gamma^{2-loop}/\Gamma^{1-loop}=\exp(8\pi k t)
$
which, even for $|k|$ as small as $5\%$ gives some three orders of magnitude uncertainty in the prediction. Hence, any potentially ``robust'' proton lifetime estimate requires at least a two-loop gauge running analysis.
\subsubsection{Scalar spectrum \& threshold effects in simple GUTs}
A consistent two-loop approach,  however, requires much more information than that needed at one loop. In particular, threshold effects should be taken into account not only around the electroweak scale but, in particular, at  the GUT scale where usually even more fields decouple. Obviously, this is impossible if the details of the high-energy spectrum are unknown or  ignored; thus, a dedicated analysis of the scalar potential is a vital ingredient here. The mediator masses can then, in principle, be determined as complicated functions of the fundamental couplings of the theory.  
\subsubsection{Planck-scale effects in the determination of the BLNV mediator masses}
Unfortunately, even this does not need to be enough for keeping the theoretical uncertainties in the mediator masses under control.  
\paragraph{Gravity smearing effects in gauge matching:} Indeed, whenever the scalar multiplet $\Phi$ responsible for the GUT-scale symmetry breaking appears in the symmetric product of two adjoint representations the gauge symmetry is not strong enough to tame the so called ``gravity smearing effects'' in the gauge coupling matching conditions~\cite{Hill:1983xh,Shafi:1983gz,Calmet:2008df}. These, at the lowest order, are related to the $d=5$ effective operator
\begin{equation}
{\cal L}^{(5)}\ni \frac{\omega}{M_{Pl}}F^{\mu\nu}\Phi F_{\mu\nu}
\end{equation}
where $F^{\mu\nu}$ is the gauge kinetic form and $\omega$ is assumed to be an ${\cal O}(1)$ coupling. This, in the broken phase, gives rise to non-universal and uncontrolled shifts in the normalization of the gauge kinetic terms which, in turn, contribute to the matching conditions for the gauge couplings, thus inflicting potentially large uncertainty in the GUT-scale determination (i.e., in the calculated masses of the relevant BLNV mediators). Let us note that, in size, this effect may be comparable to the difference between the one- and two-loop running discussed above, cf. equation~(\ref{effect}) and, hence, can not be ignored at higher than the LO level. 
\paragraph{The issue of the effective cut-off in GUTs:}
On top of that, it is often claimed that these effects may be further boosted in models with a large number of degrees of freedom which tend to lower the effective cut-off scale ($M_{Pl}$ in the formula above), see, e.g.,  \cite{Larsen:1995ax,Veneziano:2001ah,Dvali:2007hz} by as much as a factor of $[1+(N_{0}+N_{1/2}-4N_{1})/12\pi]^{-1/2}$ where $N_{0}$, $N_{1/2}$ and $N_{1}$ are the numbers of scalars, fermions and vectors, respectively. Note that this may be a particularly serious issue in supersymmetry where even the simplest models usually contain many hundreds of scalar and fermionic degrees of freedom; non-SUSY GUT tend to be in a much better shape here.   
\subsection{Flavor structure of the B\&L violating currents}
Another extremely important source of theoretical uncertainties in the proton lifetime estimates is the a-priori unknown flavour structure of the underlying GUTs. Let us simplify the situation by considering only the vector-boson-mediated $d=6$ operators of the kind a) in Figure~\ref{d6operator}; this is usually a good approximation because the Yukawa interactions of the scalar coloured triplet(s) typically emerge from the same source as the SM Yukawa interactions and, hence, share many features of their flavour structure. Among these the most important is the likely suppression of the light quark couplings in hierarchical fits, see for instance~\cite{Joshipura:2011nn,Dueck:2013gca} and references therein.

The BLNV charged-current gauge interactions, on the other hand, are governed by the unitary matrices that bring the quarks and leptons from the current to the mass bases. In the (slightly amended) notation of~ \cite{Nath:2006ut,Dorsner:2004jj} the partial proton decay widths in the prominent channels are
\begin{eqnarray}
\Gamma(p\to \pi^{0}e^{+}_{\alpha})&\propto &|c(e_{\alpha},d^{C})|^{2}+|c(e^{C}_{\alpha},d)|^{2}\,,\quad
\Gamma(p\to K^{0}e^{+}_{\alpha})\propto |c(e_{\alpha},s^{C})|^{2}+|c(e^{C}_{\alpha},s)|^{2}\,,\label{Gammas}\\
\Gamma(p\to \pi^{+}\overline{\nu})&\propto &\sum_{\alpha=1}^{3}|c(\nu_{\alpha},d,d^{C})|^{2}\,, \quad
\Gamma(p\to K^{+}\overline{\nu})\propto \sum_{\alpha=1}^{3}|B_{1}c(\nu_{\alpha},d,s^{C})+B_{2}c(\nu_{\alpha},s,d^{C})|^{2}\,,\nonumber
\end{eqnarray}
where $\alpha$ is a leptonic flavour index, $B_{1,2}$ are hadronic coefficients calculable by chiral methods, cf. \cite{Nath:2006ut}, and the $c$ factors are combinations of the quark and lepton sector diagonalization matrices
\begin{eqnarray}
M_{u}^{diag}=U_{C}^{T}M_{u}U\,,\quad M_{d}^{diag}=D_{C}^{T}M_{d}D\,,\quad M_{e}^{diag}=E_{C}^{T}M_{e}E\,,\quad M_{\nu}^{diag}=N_{C}^{T}M_{\nu}N\label{matrices}
\end{eqnarray}
defined as
\begin{eqnarray}
c(e_{\alpha},d^{C}_{\beta})&=&k_{1}^{2}(U_{C}^{\dagger}U)_{11}(D_{C}^{\dagger}E)_{\beta\alpha}+k_{2}^{2}(D_{C}^{\dagger}U)_{\beta 1}(U_{C}^{\dagger}E)_{1\alpha}\,,\nonumber\\
c(e^{C}_{\alpha},d_{\beta})&=&k_{1}^{2}\left[(U_{C}^{\dagger}U)_{11}(E_{C}^{\dagger}D)_{\alpha\beta}+U_{C}^{\dagger}D)_{1\beta }(E_{C}^{\dagger}U)_{\alpha 1}\right]\,,\label{ccoeffs}\\
c(\nu_{\alpha},d_{\beta},d^{C}_{\gamma})&=&k_{1}^{2}(U_{C}^{\dagger}D)_{1\beta}(D_{C}^{\dagger}N)_{\gamma\alpha}+k_{2}^{2}(D_{C}^{\dagger}D)_{\gamma\beta}(U_{C}^{\dagger}N)_{1\alpha}\,,\nonumber
\end{eqnarray}
where $\beta$ and $\gamma$ are quark flavour indices, $k_{1}$ and $k_{2}$ are weights inversely proportional to the masses of the GUT-scale BLNV vector mediators with quantum numbers $(\overline{3},2,+\frac{5}{6})+h.c.$ and $(\overline{3},2,-\frac{1}{6})+h.c.$, respectively.
\subsubsection{Theoretical uncertainties in the Yukawa sector fits}
Hence, in order to get a good grip on the widths~(\ref{Gammas}) one has to get as much information as possible about the diagonalization matrices in~(\ref{ccoeffs}); this can be done by matching the GUT-scale mass matrices~(\ref{matrices}) to the low-energy flavour data, i.e., quark and lepton masses and mixing parameters. This, however, is a formidable task not only due to the high level of non-linearity involved but, mainly, due to the limited amount of independent observables available. Hence, the mixing matrices in~(\ref{ccoeffs}) can not be fully determined which, in general, provides a huge and irreducible source of theoretical uncertainties in proton lifetime estimates. 

However, in some rather specific cases there may be a simple way out of this conundrum. The key point is is that the $c$-functions~(\ref{ccoeffs}) always depend on {\em products of pairs} of the flavour rotations which, sometimes, may be much better constrained by the structure of the particular model under consideration than the individual factors. For example:  
\begin{enumerate}
\item For a symmetric up-quark mass matrix and $k_{2}=0$ (this situation corresponds, for instance, to the standard Georgi-Glashow SU(5) model~\cite{Georgi:1974sy}) one has 
\begin{eqnarray}
\Gamma(p\to \pi^{+}
\overline{\nu})&\propto & k_{1}^{4}|(V_{CKM})_{11}|^{2}\,,\\
\Gamma(p\to K^{+}
\overline{\nu})&\propto & k_{1}^{4}\left[B_{1}^{2}|(V_{CKM})_{12}|^{2}+B_{2}^{2}|(V_{CKM})_{11}|^{2}\right]\,,
\nonumber\end{eqnarray}
where $V_{CKM}$ is the notorious Cabibbo-Kobayaski-Maskawa matrix.
\item For symmetric quark-sector matrices (like in renormalizable SO(10) GUTs with only $10$'s and $\overline{126}$'s active in the Yukawa sector) the formulae~(\ref{Gammas}) reduce to 
\begin{eqnarray}
\Gamma(p\to \pi^{+}
\overline{\nu})&\propto & k_{1}^{4}|(V_{CKM})_{11}|^{2}+k_{2}^{4}+2k_{1}^{2}k_{2}^{2}|(V_{CKM})_{11}|^{2}\,,\label{SO10formulae}\\
\Gamma(p\to K^{+}
\overline{\nu})&\propto & k_{1}^{4}\left[B_{1}^{2}|(V_{CKM})_{12}|^{2}+B_{2}^{2}|(V_{CKM})_{11}|^{2}\right]\,.\nonumber\end{eqnarray}
\item In the minimal renormalizable flipped SU(5), cf.~\cite{Dorsner:2004xx,Rodriguez:2013rma}, one has $k_{1}=0$ and a symmetric down-type quark mass matrix and, hence
\begin{eqnarray}
\Gamma(p\to K^+\overline{\nu})&=&0\,, \quad
\Gamma(p\to \pi^+\overline{\nu})\propto k_{2}^{4}\,. 
\end{eqnarray}
\end{enumerate}
These limits are particularly interesting because the formulae above depend (besides the masses of the BLNV vector mediators discussed in the previous section) only on the flavour information accessible at the electroweak scale. Hence, these specific partial widths are to a large degree insensitive to the details/degeneracies of the complete flavour fits! Nevertheless, all that has been said so far corresponds to the ``ideal'' situation in which all effects of the possible higher-dimensional operators are assumed to be negligible.

\subsubsection{Planck-scale effects in flavour fits}
 \paragraph{Effects in matching.}
Indeed, naively, one would expect that the per-mile gravity effects in the GUT-scale matching conditions for the effective Yukawa couplings would play a subleading role and, thus, could be neglected in flavour fits in practice. Unfortunately, due to the large mass hierarchy among the three generations of quarks and leptons the rotations~(\ref{matrices}) may be still affected considerably, especially in the most sensitive 1-2 sector. Hence, the Planck-induced higher order corrections to the Yukawa matching conditions must be taken very seriously in assessing the theoretical uncertainties in the observables~(\ref{Gammas}).     
\paragraph{The fate of the ``robust features''.} There is yet more to this: beyond the renormalizable level, even the simple qualitative conditions (e.g., the symmetry of the mass matrices)  that lead to the three particularly robust  predictions may no longer be fulfilled. If, for instance, an effective three-index tensor transforming like $120$ of SO(10) appears as a composite of other scalar fields in the model of the kind considered in case 2. above it will give rise to a (small) antisymmetric contribution to the matching conditions between the effective SM Yukawa matrices governing the masses~(\ref{matrices}) and those of the full theory which will ruin the cancellations necessary to arrive to the simplified and UV-robust formulae above. 

To this end, one may perhaps speculate that the gravitational interactions are such that these terms will actually to a large degree respect the flavour symmetry of the renormalizable sector (this may be viewed as an extreme variant of the popular minimal flavour violation (MFV) hypothesis stipulating that the Yukawas in the renormalizable Lagrangian are the only sources of flavour violation around); indeed,    
global symmetries in gravity are likely to be killed by non-perturbative effects so one may hope that, for instance, the amount of asymmetry in the ``true'' mass matrices will be so small that,  with a very good accuracy,  the ``robust features'' of the specific scenarios above will be retained. 
 
\section{THE MINIMAL RENORMALIZABLE SO(10) GUT} 
From what has been said one may conclude that it is virtually impossible to provide a theoretically robust NLO prediction for the proton lifetime in the classical  GUT framework as the proximity of the Planck scale and the size of its effects on the vital ingredients of these calculations makes the theoretical uncertainties way larger than the desired order-of-magnitude ballpark. Nevertheless, there is a very special setting in which the leading order gravity effects may remain under control to such a degree that it is at least worth trying to perform an NLO analysis within. 

Indeed, the renormalizable non-supersymmetric SO(10) model in which the GUT-scale gauge symmetry is broken by the 45-dimensional adjoint scalar representation and the Yukawa couplings are governed by the SO(10) vector(s) and 5-index antisymmetric tensor(s) (i.e., $10$ and $\overline{126}$), see, e.g.,~\cite{Chang:1984qr,Deshpande:1992au}, may overcome the main issues discussed above. Due to the antisymmetry of $45$ the leading order gravity smearing effects in the gauge matching are absent and, hence, the masses of the BLNV mediators may be, in principle, determined to a sufficient accuracy. At the same time, the symmetric nature of all the Yukawa couplings at play may justify the use of the formulae~(\ref{SO10formulae}) and, thus, overcome also the issue with the limited amount of the flavour information available at the low scale. 

Interestingly enough, this very special and beautiful model has been ignored for more than two decades due to the peculiar tachyonic instabilities revealed in its scalar sector back in 1980's~\cite{Yasue:1980fy,Anastaze:1983zk,Babu:1984mz}; only recently it has been shown~\cite{Bertolini:2009es} that this is a mere artefact  of the tree-level approximations used therein and that the model makes perfect sense as a truly quantum theory. Since then, it has been a subject of several dedicated analyses~\cite{Bertolini:2013vta,Kolesova:2014mfa,Bertolini:2012im} which will hopefully culminate into a complete and robust NLO prediction in a not-so-distant future.   
\section{CONCLUSIONS AND OUTLOOK} 
Given its enormous scientific potential the next generation of the neutrino oscillation and proton decay machines clearly deserves the highest level of support the HEP community can generate. To that end, a robust NLO calculation of the proton lifetime would be highly desirable as it may contribute significantly to making a clear case for such a huge investment. Unfortunately, there are several sources of almost irreducible theoretical uncertainties that, so far, made this exercise virtually intractable. Nevertheless, after more than 20 years in a limbo, the minimal renormalizable SO(10) grand unified model in its purely quantum version seems to be the candidate framework in which this dream may finally come true.     

\section{ACKNOWLEDGMENTS}
The work of M.M. is supported by the Marie-Curie Career Integration Grant within the 7th European Community Framework Programme
FP7-PEOPLE-2011-CIG, contract number PCIG10-GA-2011-303565. The work of H.K. is supported by the Grant Agency of the Czech Technical University in Prague, grant No. SGS13/217/OHK4/3T/14. The work of M.M. and T.M. is also supported by the Foundation for support of science and research ``Neuron''. M.M. is indebted to the organisers of the marvellous CETUP'15 for hospitality and support.

\end{document}